\documentclass[aps, prb, superscriptaddress, showpacs,twocolumn]{revtex4-1}
\usepackage{color}

\pdfoutput=1
\usepackage{graphicx}

\begin{document}

\title{Intrinsic Magnetism in Nanosheets of SnO$_{2}$: A First-principles Study}

\author{Gul Rahman }
\email{gulrahman@qau.edu.pk}
\affiliation{Department of Physics,
Quaid-i-Azam University, Islamabad 45320, Pakistan}

\author{V\'{\i}ctor M. Garc\'{\i}a-Su\'arez}
\affiliation{Departamento de F\'isica, Universidad de Oviedo, 33007 Oviedo Spain}
\affiliation{Nanomaterials and Nanotechnology Research Center (CINN), CSIC - Universidad de Oviedo, Spain}
\affiliation{Department of Physics, Lancaster University, Lancaster LA1 4YW, United Kingdom}

\author{J. M. Morbec}
\affiliation{Instituto de Ci\^encias Exatas, Universidade Federal de Alfenas, 37130-000, Alfenas, MG, Brazil}

\begin{abstract}
We propose intrinsic magnetism in nanosheets of SnO$_{2}$, based on first-principles calculations.
The electronic structure and spin density reveal that $p$ orbitals of the oxygen atoms, surrounding Sn vacancies, have a non itinerant nature which gives birth to localized magnetism. A giant decrease in defect formation energies of Sn vacancies in nanosheets is observed. We, therefore, believe that native defects can be stabilized without any chemical doping. Nanosheets of different thicknesses are also studied, and it is found that it is easier to create vacancies, which are magnetic, at the surface of the sheets. SnO$_{2}$ nanosheets can, therefore, open new opportunities in the field of spintronics.
\end{abstract}

\maketitle
The quest for room temperature (RT) ferromagnetism in diluted magnetic semiconductor~(DMS), where transition metal (TM) (magnetic) impurities are doped into semiconductor hosts, e.g., Mn in GaAs~\cite{Ohno}, has motivated many theoretical and experimental studies. TM impurities were found however to produce intrinsic defects which limit its application in the field of spintronics at RT. Next ideal candidates for RT applications in spintronics are oxide-based DMS systems, which not only have high $T_\mathrm{C}$, but also large magnetic moments~\cite{Ogale}. Such oxide-based DMS also faced challenges, i.e., the determination of the origin of their magnetic properties and the control of these properties in oxides. The difficulties arise from the presence of defects (cation/anion vacancies), which can be beneficial for magnetism but whose degree of influence is not completely understood. It is believed that some oxides show magnetism without any magnetic impurities~\cite{Rahman2008, HfO2,ZnO,CeO2,MgO} mainly due to cation vacancies. The role of vacancies (either cation or anion) in magnetism can not then be ignored~\cite{Chang}.

For the last few years extensive work (both experimental and theoretical) has been done to understand the origin of magnetism in materials with defects. Some of these new discoveries also made possible to have magnetism in nonmagnetic insulators/semiconductors doped with light elements~\cite{Rahman2010,NSnO2,CuSnO2}.
Very recently, some of us found that C can also induce magnetism in SnO$_{2}~$\cite{Rahman2010}, and later on this new theoretical idea was shown experimentally
by Hong \textit{et al}~\cite{Hong}. Therefore, it is now believed that magnetism can exist either in materials with vacancy or light elements-doped systems~\cite{Rahman2008, HfO2, ZnO,CeO2,MgO,Rahman2010,NSnO2,CuSnO2}. Usually, cation vacancies, which are a source of magnetic moments, have large formation energies and it is not easy to produce them experimentally. Therefore, there remains an open and challenging question how intrinsic defects (vacancies) can be stabilized in nonmagnetic materials. Here, we find a new way of stabilizing defects in SnO$_{2}$. {We show that intrinsic defects can be easily stabilized by reducing the dimensionality of SnO$_{2}$ }. This is a very unique way of reducing the formation energy of vacancies without any alien impurities. {Note that there are several experimental reports on the fabrication of nanostructured SnO$_{2}$~\cite{Dia,Leite,Dia2,Law,Khan}. Beltran \textit{et al.}~\cite{Beltran}, have discussed the thermodynamic stability of nanostructured SnO$_{2}$ and have shown (both theoretically and experimentally) the possibility of nanosheets formation.}

\begin{figure}[!h]
\includegraphics[width=0.4\textwidth, angle=90]{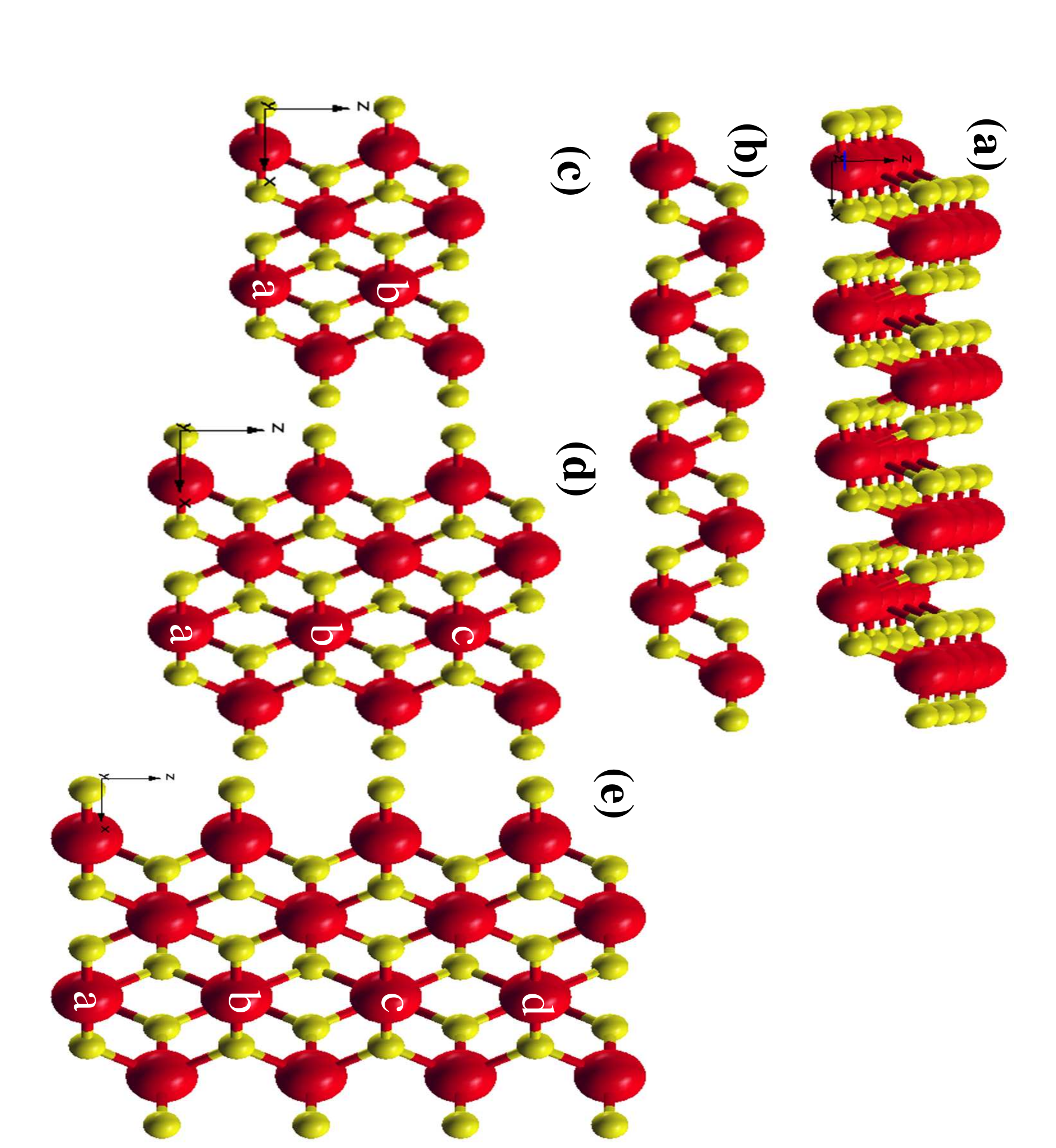}
\caption{(Color online) Structure of nanosheets of SnO$_{2}$.
(a) shows a $4\times4\times1$ nanosheet.
(b) shows the side view of (a).
(c), (d), and (e) show $2\times2\times2$, $2\times2\times3$, and $2\times2\times4$ nanosheets, respectively.
Labels $a,...,d$ indicate the location of the Sn vacancy.}
\label{sheet_fig}
\end{figure}

First-principles calculations based on density functional theory within the local spin density approximation (LSDA)~\cite{lsda} and generalized gradient approximation (GGA)~\cite{gga} were performed using the {SIESTA} code,~\cite{siesta} which employs norm-conserving pseudopotentials and linear combinations of atomic orbitals. The convergence of the basis set (a double-zeta polarized, DZP) and other computational parameters (real space energy cutoff, 200 Ry; $k$-point sampling, 9x9x1 $k$-points) were carefully checked. All atomic positions were fully relaxed until all atomic forces were smaller than 0.05 eV/\AA, which proved enough to converge the structure of the system.~\cite{Rahman2008}
Different single-layer nanosheets [$2\times2\times1$ (24 atoms), $3\times3\times1$ (54 atoms) and $4\times4\times1$ (96 atoms)]
were considered to take into account the
effect of the vacancy concentration
on the electronic and magnetic properties
(a $4\times4\times1$ nanosheet is shown in Fig.~\ref{sheet_fig}(a) and (b)). We also considered nanosheets of different thicknesses [$2\times2\times n$, where $n=2,3,4$],
as depicted in Fig.~\ref{sheet_fig},
to investigate the influence of the position of Sn vacancy on the defect formation energies.
{The Sn and O vacancies were simulated
within the supercell approach with periodic boundary
conditions in the plane and open boundary conditions perpendicular to the plane;
a vacuum region of 10\,\AA \ along the direction normal to the surface of the nanosheet was found enough to achieve convergence on this parameter.} 

\begin{figure}[!h]
\includegraphics[width=0.22\textwidth]{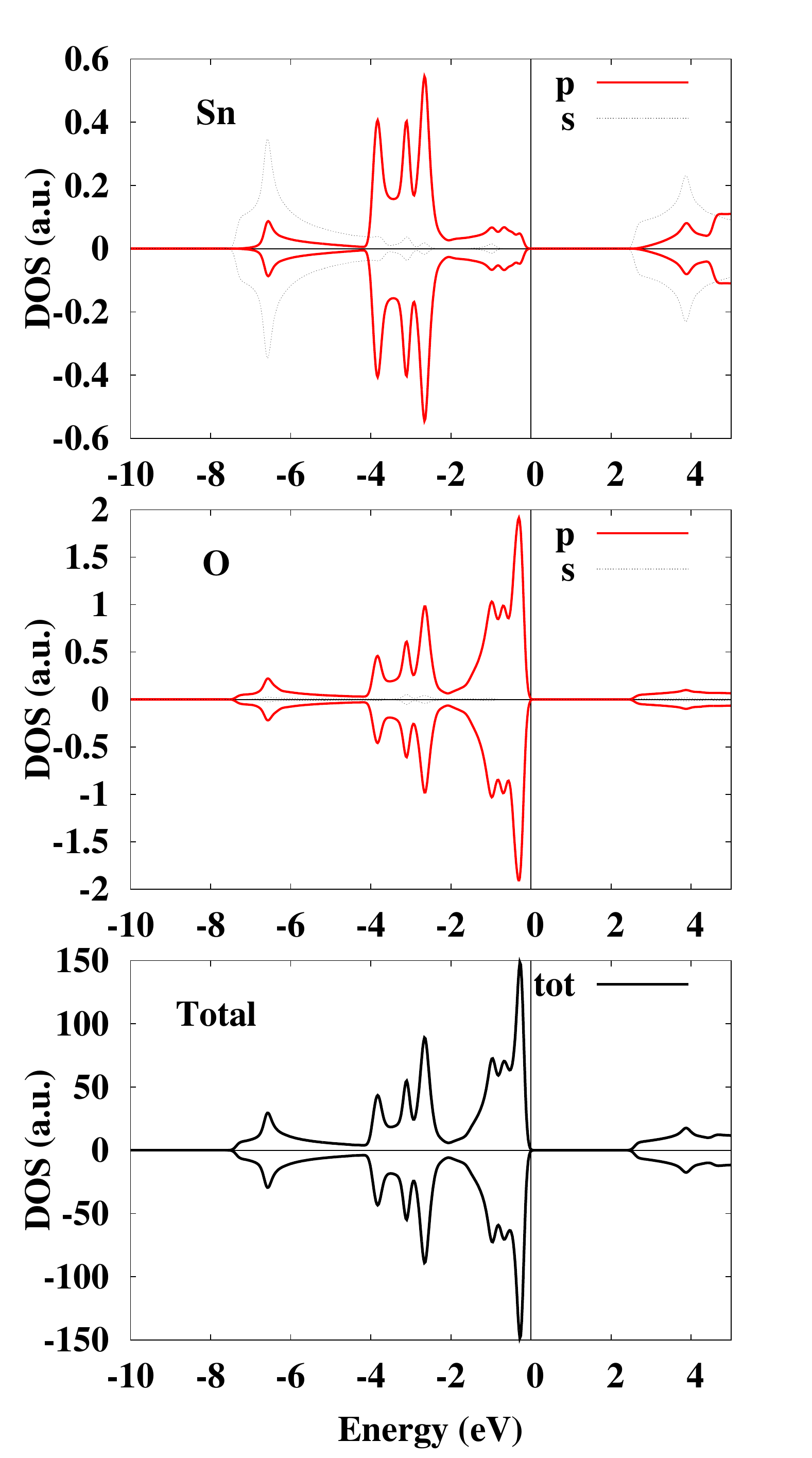}
\includegraphics[width=0.22\textwidth]{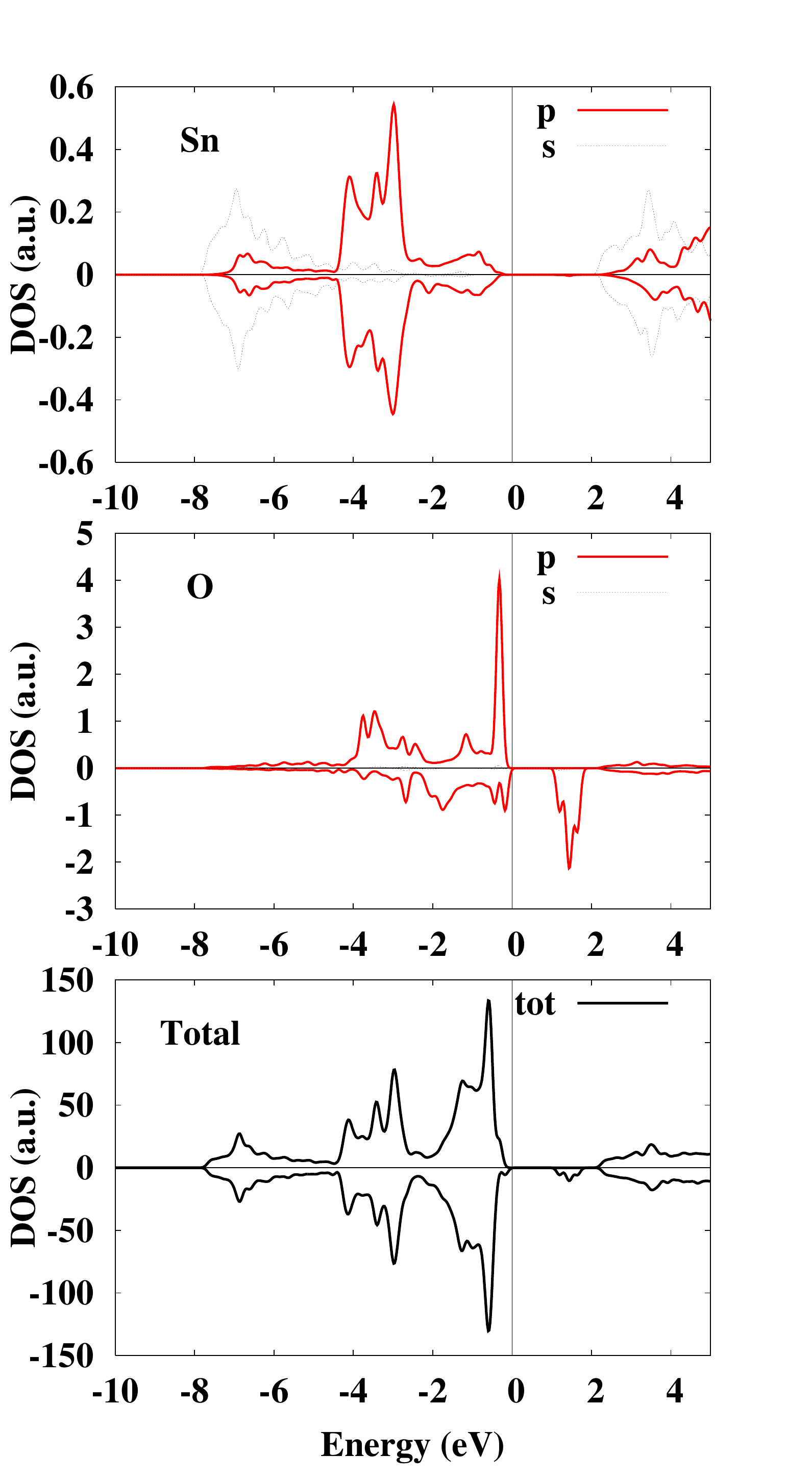}
\caption{(Color online)The total and projected density of states (PDOS) of pure (a)(left-side) and with a single Sn vacancy (b)(right-side) SnO$_{2}$ sheets. Solid and dashed lines show $p$ and $s$ orbitals of O/Sn atoms. Thick solid lines in the bottom panel show the total density of states. The Fermi energy ($E_\mathrm{F}$) is set to zero. }
\label{PureDOS}
\end{figure}

First, we carried out nonmagnetic~(NM) and magnetic (M) calculations on pristine SnO$_{2}$ nanosheets  and we found no trace of magnetism. The total and atom-projected density of states (DOS and PDOS) are shown in Fig.~\ref{PureDOS}(a). It is clear that the pristine sheet is a wide band gap nonmagnetic material where all bands below the Fermi level ($E_\mathrm{F}$) are completely occupied, similar to bulk SnO$_{2}$. The DOS near the Fermi level  is dominated by the $p$ orbitals of O in the valence band and the Sn orbitals (with some weight of the $p$ O) in the conduction band.
Once confirmed that pure SnO$_{2}$ nanosheets are nonmagnetic,
we carried out simulations of native defects (i.e., Sn and O vacancies) in a single-layer SnO$_{2}$ nanosheets.
We must note that in these nanosheets Sn is not surrounded by six  but by four oxygen atoms (see Fig.~\ref{sheet_fig}(b)). Then dangling bonds are expected in such structures. We created the Sn and O vacancies at the center of the sheets. Interestingly, we found that Sn vacancies induce  very large magnetic moments of $4.0\mu_\mathrm{B}$. However, the O vacancies do not induce magnetism, this behavior is similar to bulk SnO$_{2}$~\cite{Rahman2008}.  A detailed Mulliken analysis shows that the magnetic moment is mainly contributed by the oxygen atoms surrounding the Sn vacancy. Each of the surrounding four O atoms carry a local magnetic moment of about $1.0\mu_\mathrm{B}$. This can be easily understood by taking into account that the Sn vacancy creates localized dangling bonds at the O sites which give rise to local magnetic moments following Hund's rules~\cite{blundel}. The magnetic coupling (ferromagnetic or antiferromagnetic) between the O atoms depends on the relative orientation and distance between the cation vacancies~\cite{Rahman2008,Zunger}.   The localization of the O magnetic moment is further confirmed by LDOS and spin-density maps.

Figure~\ref{PureDOS}(b) shows the calculated total DOS and PDOS of a SnO$_{2}$ sheet with a single Sn vacancy. We only show the PDOS of one of the oxygen atoms, since the remaining three oxygen atoms around the vacancy displayed  
the same behavior. The other oxygen atoms also contribute to the total DOS but their contributions are very small as compared with the oxygen atoms surrounding the Sn vacancy.  The total DOS shows that the Sn vacancy induces magnetism in SnO$_{2}$ nanosheets but does not destroy the insulating behavior of SnO$_{2}$. Therefore, SnO$_{2}$ sheets with Sn vacancies can be assigned to the class of magnetic insulators. It is clear from this figure that the magnetization is mainly contributed by the oxygen atoms. The oxygen spin-up $p$ band is completely occupied whereas the spin-down band has a smaller occupation below $E_\mathrm{F}$ and some weight above it. The electronic structure near the Fermi level is dominated by the oxygen $p$ orbitals, but there is also a small contribution from the $p$ and $s$ orbitals of Sn below the Fermi level, where oxygen states hybridize with the Sn states. This partial/complete occupation of the oxygen $p$ orbitals illustrates their localized nature. It must be noted that such localized $p$ states are absent in bulk SnO$_{2}$~\cite{Rahman2008}, which suggest that this localized behavior of the $p$ orbitals can be attributed to the reduced dimensionality of SnO$_{2}$.
\begin{figure}
\includegraphics[width=0.25\textwidth, angle=90]{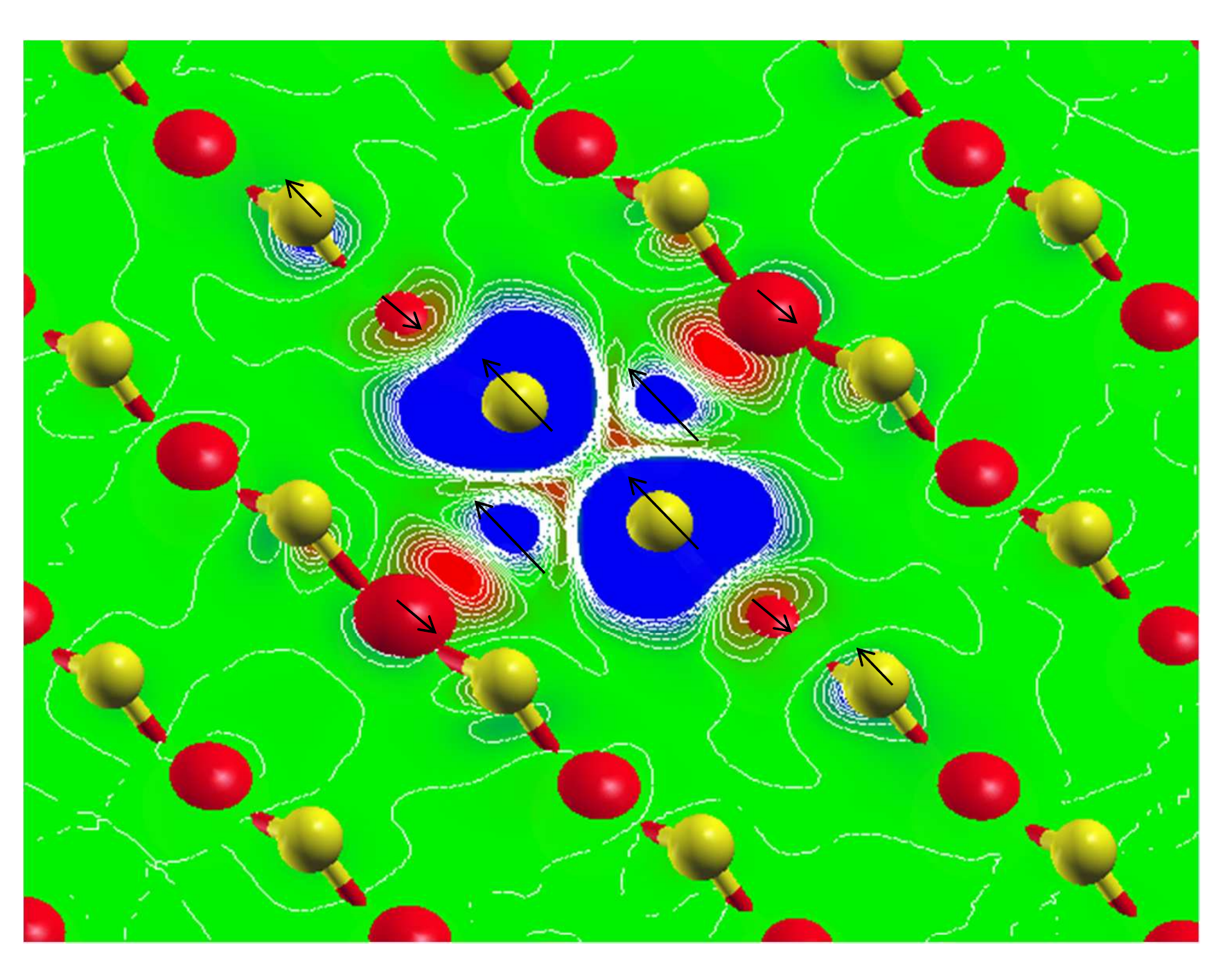}
\caption{(Color online) Spin density of a single sheet SnO$_{2}$ with a single Sn vacancy. Red (yellow) balls show the tin (oxygen) atoms. Arrows represent the direction of the spin magnetic moments.  Blue (red) colors in the plane show negative (positive) spin polarization of the atoms.}
\label{spindos}
\end{figure}

To shed more light on the nature of bonding and the source of magnetism in SnO$_{2}$ nanosheets, we calculated the spin-density on a plane that contains the Sn vacancy (see Fig.~\ref{spindos}). As can be seen, the spin-density is strongly localized at the oxygen atoms, which couple ferromagnetically. The spin-density is, therefore, concentrated around the Sn vacancy and does not extend to other O atoms, which signals the non-itinerant behavior of the oxygen $p$ electrons. There are also small negative induced magnetizations at the nearest Sn atomic sites. This shows that the negative polarization of the Sn atoms is essential for mediating magnetism between oxygens in the context of superexchange interactions~\cite{blundel}, which play a decisive role in governing localized magnetism. We repeated the same calculations for different sheets (i.e., varying the O or Sn vacancy concentrations), and we found the same conclusion, i.e., the Sn vacancy induces magnetism, which is mainly contributed by the surrounding oxygen atoms. The magnetic  moments are summarized in Table~\ref{tableSnO2}.
\begin{table}
\caption{Calculated formation energies (in units of eV) in equilibrium (Eq.), Sn-rich, and O-rich conditions for a Sn vacancy in SnO$_{2}$ sheets as a function of Sn vacancy concentration (in \%). The last column shows the total magnetic moment (MM) per unit cell (in units of $\mu_\mathrm{B}$).}
\begin{center}
\begin{tabular}{cccccccccccc}
\hline \hline
Sn vacancy(in \%) &Eq.& $\;\;\;$Sn-{rich}$\;\;\;$&O-{rich} & $\;\;\;$Total MM\\
\hline
12.500&1.72&	$9.29$	&$1.72$	&$4.00$\\
$\;\;6.250$&1.76&	$9.32$	&$1.76$	&$4.00$\\
$\;\;3.125$&1.80&	$9.37$	&$1.80$	&$4.00$\\
\hline\hline
\end{tabular}
\label{tableSnO2}
\end{center}
\end{table}

In order to determine the experimental feasibility of such configurations it is essential to determine the defect formation energy $E_\mathrm{f}$. We calculated the formation energies of the Sn and O vacancies in these nanosheets using the formula
\begin{displaymath} {E_\mathrm{f}=E_\mathrm{D}(\mathrm{SnO}_{2})-E_\mathrm{P}
(\mathrm{SnO}_{2})+n\mu_\mathrm{X}},
\end{displaymath}
where $E_\mathrm{P}$ ($E_\mathrm{D}$) is the total energy of pristine (defect, containing either Sn or O vacancy) sheets of SnO$_{2}$, $\mu_\mathrm{X}$ is the chemical potential of X(X $=$ Sn or O), and $n$ is number of X atoms removed. In equilibrium conditions, the chemical potentials of O and Sn satisfy the relationship
$\mathrm{\mu_{Sn} + 2\mu_O = \mu_{SnO_{2}}} $, where
$\mu_\mathrm{SnO_{2}}$, the chemical potential of bulk SnO$_2$, is a constant value calculated as the total energy per SnO$_2$ formula unit.
Under Sn rich conditions,
$\mu_\mathrm{Sn}$=E(Sn$^{\mathrm{metal}}$) and
$\mu_\mathrm{O}=\left(\mu_\mathrm{SnO2}-\mu_\mathrm{Sn} \over 2\right)$, whereas in O rich conditions, $\mu_\mathrm{O}$ = $\left(1\over 2\right)$E(O$_2$) and
$\mu_\mathrm{Sn}=\mu_\mathrm{SnO2}-2\mu_\mathrm{O}$.

The calculated $E_\mathrm{f}$ as a function of Sn vacancies, in equilibrium, Sn-rich and, O-rich conditions, is summarized in Table~\ref{tableSnO2}. As can be seen, the formation of Sn vacancies is favored under O-rich conditions, and it decreases as the concentration of Sn vacancies increases in the SnO$_{2}$ nanosheets. It is very encouraging to see that $E_\mathrm{f}$ is significantly decreased in nanosheets as compared with $E_\mathrm{f}$ in bulk SnO$_{2}$, which is $\sim 6.88$ and $14.44\,${eV} in O-rich and Sn-rich conditions, respectively. The smaller values of $E_\mathrm{f}$ in nanosheets of SnO$_{2}$ ($\sim 75\%$ smaller than the bulk values for the O-rich case) further guarantee the experimental realization of our proposed systems. We must also mention that previously synthesized samples of transition metal doped SnO$_{2}$ have much larger $E_\mathrm{f}$, e.g., $7.0,\,2.0$ and $3.27\,$ in O-rich conditions for Zn-, Cr-, and Co-doped SnO$_{2}$, respectively~\cite{Zn,Cr,Co}. This further makes us confident about the experimental verification of our proposed systems. On the other hand, $E_\mathrm{f}$ of O vacancies in SnO$_{2}$ nanosheets is calculated to be $\sim-0.50$ and $\sim 3.30\,${eV} in Sn- and O-rich conditions, respectively. Although $E_\mathrm{f}$ is still smaller for O-vacancies in nanosheets as compared to the bulk, we believe these vacancies do not play a significant role in SnO$_{2}$ in terms of $d^{0}$ magnetism, since they do not induce magnetic moments. However, we can not forget the role of O vacancies in promoting magnetism in TM-doped SnO$_{2}$~\cite{Cr,Co}.
\begin{figure}
\includegraphics[width=0.3\textwidth, height=0.6\textwidth]{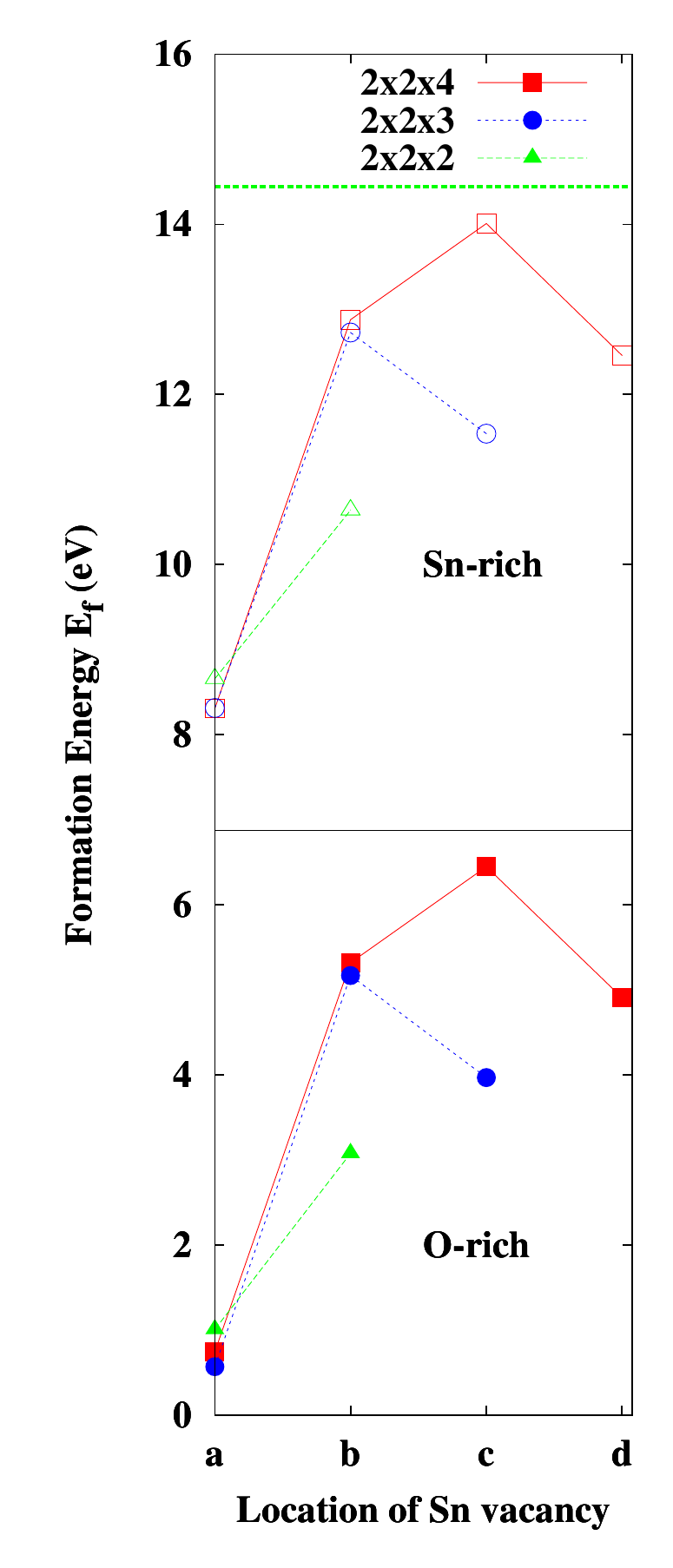}
\caption{(Color online) Calculated defect (V$_\mathrm{Sn}$) formation energies $E_\mathrm{f}\;$ (eV) for SnO$_{2}$ nanosheets with different thicknesses. The $x$-axis shows the location of V$_\mathrm{Sn}$ in different sheets as marked in Fig~\ref{sheet_fig}. Square, circle, and triangle represent $E_\mathrm{f}$ of $2\times2\times4$, $2\times2\times3$, and $2\times2\times2$ thick sheets of SnO$_{2}$, respectively. Filled (empty) symbols represent $E_\mathrm{f}$ in O-rich (Sn-rich) conditions.  The horizontal dashed (solid) line shows $E_\mathrm{f}$ of bulk SnO$_{2}$ in Sn-rich (O-rich) conditions.}
\label{formation}
\end{figure}

Next, we considered nanosheets of different thicknesses
[$2\times2\times2$, $2\times2\times3$, and $2\times2\times4$]
 and created Sn vacancies in different layers, as shown in Fig.~\ref{sheet_fig}(c)-(e). The calculated formation energies are summarized in Fig.~\ref{formation}. The Sn vacancy created at the surface of the sheets [marked as \textbf{a} in Fig.~\ref{sheet_fig}(c)-(e)])
has the lowest formation energy, $\sim 0.70\,${eV}, which is a giant decrease in $E_\mathrm{f}$ as compared with the bulk value. Such a small value further guarantees that Sn vacancies will be very active at the surface of the sheets and will easily be achieved experimentally. However, as the defect was moved to the second/third layer of the sheets
[for example, sites \textbf{b, c} in Fig.~\ref{sheet_fig}(e)],
$E_\mathrm{f}$ increased and approached the bulk value in both Sn-rich and O-rich conditions. This behavior can be understood by taking into account bond saturation, which increases the stabilization of each atom inside the sheet. However, near the surface layers, the broken Sn-O bonds reduce $E_\mathrm{f}$ significantly. In any case, whatever the location of the Sn vacancy in these sheets, $E_\mathrm{f}$ is always smaller than the bulk values. {Once we have calculated $E_\mathrm{f}$ of the point defects, its concentration ($c$), which depends on its formation energy, in thermodynamic equilibrium is given by
$c=n_{\mathrm{site}}\,\mathrm{exp}  \left(\frac{-E_\mathrm{f}}{k_\mathrm{B}T}\right)$.
Where $E_\mathrm{f}$ is the defect formation energy, $n_\mathrm{site}$ is the number of sites the defect can be incorporated on, $k_{\mathrm{B}}$ is the Boltzmann constant,
and $T$ is the temperature. The above expression shows that defects
with high formation energies will occur in low concentrations.} It is also very interesting that the magnetic moment ($4.0\mu_{B}$) remains robust when the thickness of the sheets changes or the Sn vacancy moves.%
We found, however, a significant increase in the local magnetic moments of the oxygen atoms at the surface of the sheets. The two surface O atoms have a local magnetic moment $\sim 1.25\mu_{B}$, which is almost independent of the sheet thickness. {After relaxing the structure, the bond lengths of the Sn and O atoms surrounding the Sn vacancy were increased by $\sim 0.07°$\AA. This relaxation was anisotropic, since the in-plane and out-of-plane bond lengths were slightly different. The structural relaxation had no significant effect on the total magnetic moments. However, a slight change in the local magnetic moments of the Sn and O atoms was observed.}

{Our extensive calculations clearly demonstrate the possibility of magnetism without impurities in nanosheets of SnO$_{2}$. This possibility is proved by the fact that the defect formation energies of the Sn vacancies are much smaller than those in bulk SnO$_{2}$. This makes us believe that such magnetic nanosheets are experimentally possible. Interestingly,  Wang \textit{et al}. \cite{wang, wang2} recently synthesized nanosheets of SnO$_{2}$ and showed room temperature ferromagnetism in fresh and annealed systems. Such experimental report further confirms that room temperature ferromagnetism in nanosheets of SnO$_{2}$ is enhanced by the small values of the defect formation energies.}
{Finally, test calculations, using GGA and LDA+$U$, were also carried out on pristine nanosheets and sheets with Sn vacancies. We found the same behavior, i.e., pristine nanosheets are nonmagnetic and Sn vacancies in nanosheets induce magnetism. Note that previous theoretical calculations have shown that the inclusion of the $U$ term does not change the magnetic moments caused by vacancies~\cite{Sanvito,Fernandes}. Such term will have some effect on the electronic density of states~\cite{Sanvito2}, but it will not change the main conclusion drawn here. Note that we focused on the origin of magnetism in SnO$_{2}$ nanosheets and our calculations agree with the recent experimental data~\cite{wang, wang2}. The macroscopic magnetic coupling (FM or AFM) between Sn vacancies would required much larger simulations, which is beyond the scope of the present study. }

In summary, first-principles calculations based on density functional theory within local density approximation have been carried out to investigate magnetism
and formation energies of native defects in SnO$_{2}$ nanosheets with different thicknesses.
We showed that Sn vacancies induce large magnetic moments and O vacancies do not induce magnetism.
The origin of localized magnetism was discussed using electronic densities of states and spin density contours.
The defect formation energies for different thicknesses were also calculated and it was found that Sn vacancies need much smaller energies in nanosheets than in bulk SnO$_{2}$. These results indicate that it is easy to create Sn vacancies at the surfaces of SnO$_{2}$ nanosheets. We conclude that intrinsics defects can be easily stabilized in nanosheets of SnO$_{2}$ without any alien impurities.

GR acknowledges the cluster facilities of NCP, Pakistan.
VMGS thanks the Spanish Ministerio de Ciencia e Innovaci\'on for a Ram\'on y Cajal fellowship (RYC-2010-06053).
JMM acknowledges the computational facilities provided by CESUP-UFRGS, Brazil.


\end{document}